# Model Pluralism

*Walter Veit[1]*

August 2019

**Abstract**

This paper introduces and defends an account of model-based science that I dub model pluralism. I argue that despite a growing awareness in the philosophy of science literature of the multiplicity, diversity, and richness of models and modeling-practices, more radical conclusions follow from this recognition than have previously been inferred. Going against the tendency within the literature to generalize from single models, I explicate and defend the following two core theses: (i) any successful analysis of models must target sets of models, their multiplicity of functions within science, and their scientific context and history and (ii) for almost any aspect $x$ of phenomenon $y$, scientists require multiple models to achieve scientific goal $z$.

**1 Introduction**
In this paper, I introduce an account of models, modeling, and model-based science (abbreviated henceforth as MMM) that I dub *model pluralism.* Accordingly, the title of this paper is intended to reflect the incredible diversity of models, their purposes and roles in science – components of MMM, that though starting to become more and more recognized, undermine an essential tenet of the literature at large.

In the face of the incredible diversity of both models and modeling practices, a naive form of model monism appears untenable. Mitchell (2002) calls pluralism in science simply a "fact", which is no overstatement. However, there is a certain ambiguity in how model pluralism might be understood: is it (i) merely a descriptive claim, stating that models and modeling practices in science are incredibly diverse, or (ii) a prescriptive position stating that model diversity is to be sought and embraced? Despite arguing for both positions, it is useful to distinguish (i) as model diversity from (ii) i.e. model pluralism. The same ambiguity is naturally found in the

[1] Carnegie Mellon University, Pittsburgh, Pennsylvania, USA & University of Bristol, Bristol, UK

Corresponding Author: Walter Veit, Department of Social and Decision Sciences, Carnegie Mellon University, Pittsburgh, Pennsylvania, USA 15213.
Email: wrv@andrew.cmu.edu

term *scientific pluralism*. It can either be a descriptive statement about the diversity found in science, or a prescriptive position according to which this diversity is to be embraced (at least to some extent). To the extent that I see model diversity as an unavoidable feature of science, rather than a problem, I will refer to both (i) and (ii) interchangeably when speaking of *model pluralism*.

Far from universally recognized, however, the diversity of models is still seen as something to be avoided, raising more problems than it solves. In this paper, I shall argue that the inverse is true: model pluralism solves a variety of old philosophical problems surrounding the use and misuse of highly abstract and idealized models. More so, I argue that model pluralism is an unavoidable feature of model-based science and should be wholeheartedly embraced. Philosophers such as Cartwright (1983, 2009), Hausman (1992), Mäki (2009), de Donato & Zamora-Bonilla (2009); Grüne-Yanoff (2009); Knuuttila (2009), Kuorikoski & Lehtinen (2009), Northcott & Alexandrova (2015), Fumagalli (2016), and Marchionni (2017), may have been exceedingly critical of confidence in highly idealized models by solely focusing on the relationship between a single model and its relationship to the real world. Only recently have philosophers of science started to shift their focus towards the epistemic contribution of sets of models, rather than single models as such.[2] This startling omission within the literature shall be remedied here.

Accordingly, I shall argue that despite a slow transition in the MMM literature towards a growing awareness of the plurality, diversity, and richness of models and modeling practices, more radical conclusions follow from this recognition have previously been inferred. In doing so, I defend the following two core theses of model pluralism: (i) any successful analysis of models must target sets of models, their multiplicity of functions within science, and their scientific context and history and (ii) for almost any aspect $x$ of phenomenon $y$, scientists require multiple models to achieve scientific goal $z$.[3]

## 2 Four Kinds of Model Pluralism

Over the last 50 years, the philosophy of science has increasingly shifted its trajectory from a focus on theories and laws to a focus on models.[4] This shift in attention unsurprisingly coincides with the evidentially ever-growing use by scientists of the words 'model' and 'modeling' to describe their work. However, hardly any term has been more contested in the philosophy of science than the term 'model'.[5] Many philosophers of science seem to be under the impression that once we figure out what 'models' are and how they work, we can finally get a handle on all the other pressing questions in the philosophy of science. I argue that this way of thinking about models is grounded in a mistake. In this, I am echoing a worry the eminent evolutionary biologist and modeler John Maynard Smith once expressed when asked about Popper and the philosophy of science:

> more generally, my impression is that […] scientists who take the philosophy of science seriously and allow their scientific research to be influenced by philosophical preconceptions, are far more likely to do themselves harm than good. […] And I think it happens again and again in science. So, I'm... I'm a sucker here, I mean, I love reading philosophy of science, I find it interesting. I feel all ready to argue about it. But I do not believe one should allow oneself to be influenced by it, when actually thinking about science. (1997)

---

[2] See Knuuttila (2011); Muldoon (2007); Wimsatt (2007); Weisberg (2007), 2013; Ylikoski & Aydinonat (2014); Lisciandra (2017); Aydinonat (2018b); Grüne-Yanoff & Marchionni (2018).
[3] I alluded to this position in an earlier paper on the role of models in the scientific study of morality (see Veit forthcoming).
[4] This is not to deny that there have been philosophical discussions of models for over a century, see for instance Boltzmann (1902/1974).
[5] See Godfrey-Smith (2006)

These are quite radical words and though Maynard Smith's generalization is untenable, there is an important insight to be extracted. If modelers approach the literature on MMM with the goal of figuring out what models are and what a good model is, they might easily find themselves lured into naive monistic views of models that fail to represent the full epistemic richness of modeling-practices, thereby turning them into worse scientists than they could have otherwise been. This could happen in several ways. For instance, models with epistemic functions not recognized by narrow monist accounts might be discarded. Consider, for example, models with weak predictive power, but great generality and insight into complex systems.[6] Perhaps Maynard Smith's worry is most strongly illustrated in economics, where Michael Friedman (1953) and Karl Popper (1959, 1963), the latter being the explicit target of Maynard Smith, are taken to have the "last word" in any methodological debate, and are frequently invoked to both defend and criticize the work of modelers.[7]

Perhaps paradigm cases for monist thinking, Friedman (1953) argued that the only relevant desiderata in science concerns predictive power, while Popper (1959, 1963) argued that what matters is falsifiability. These accounts of science have received extensive criticism over the last decades. Indeed, Rosenberg (2009) argued that Friedman's naive picture "single-handedly justified" (p. 57) the existence of philosophy of economics as a distinct profession. Economists, however, are unlikely to read much if any work on methodology beyond Friedman and Popper.[8]

The obvious counter to this concern is, that it is not an argument against the philosophy of science, but rather *bad* philosophy of science, which limits, rather than fosters, insights into the richness of the epistemic tools available to scientists. Some debates in the philosophy of science, however, rest on such monist motivations to seek univocal accounts of laws, explanation, species, functions, preferences, and many more. As Godfrey-Smith (2003) emphasized: "[o]ne of the hazards of philosophy is the temptation to come up with theories that are too broad and sweeping" (p. 5). Accordingly, I argue that this hazard has not been avoided in the philosophical literature on MMM.

Pluralist views are becoming increasingly popular in various debates in the philosophy of science. Nevertheless, I argue that the more pluralist views emerging in the MMM literature do not go far enough. Here, Michael Weisberg's (2013) recent monograph *Simulation and Similarity* offers an elegant starting point of contrast as the arguably most comprehensive and pluralist account in the recent philosophical literature on MMM.[9] Indeed, Weisberg has perhaps done more than anyone to move us away from monist conceptions of models and his contribution must be seen as a substantial improvement upon a literature that has often only focused on one or a small set of models.[10] However, as I shall show, more radical conclusions follow from Weisberg's (2007, 2013) own arguments for "multiple-model idealization" than he himself anticipated. Model monism and model pluralism stand in an inherent tension to each other. Model monists seek one model: the best model, the perfect model, the model that is general, precise, and realistic. Some may consider this a strawman, but the manifold responses to Levins' (1966) call for multiple models would suggest otherwise.

When a scientific discipline diversifies, with more models and modeling practices emerging, model monists see this as an essential problem. They might justify the present diverse state by arguing that given the premature state of a scientific discipline, such

---

[6] Many models in the evolutionary game theory literature in both biology and the social sciences have these features.

[7] See Rosenberg (2009) and Rodrik (2015).

[8] The scope of this paper does not allow me to go deeper into the negative influence Popper and Friedman had on economics, but Rodrik (2015) provides a book-length treatment of the matter.

[9] However, Gelfert's (2016) recent monograph, *How to Do Science with Models: A Philosophical Primer*, is a contender on this position.

[10] See also O'Connor and Weatherall (2016) and Aydinonat (2018b).

diversification is the best way of hitting upon the right model. Kitcher (1991) argued that a diversity of theories could simply be a division of cognitive labour in order to find the right theory. This argument is of course readily applicable to models, but nevertheless paints diversity as something that should eventually be overcome. In contrast to model monism, which I see as a non-starter, model pluralism is an ambiguous term. In order to bring some clarity into this debate, I propose the following distinction between four different kinds of model pluralism:

| Four Kinds of Model Pluralism | Definition |
| --- | --- |
| 0. Model Monism | Search for the perfect (general) model. |
| 1. Weak Model Pluralism | There are many phenomena, and scientists need different models to explain/predict these different phenomena. |
| 2. Weakly Moderate Model Pluralism | Each phenomenon has many different aspects, and scientists need different models to explain/predict these different aspects of a single phenomenon. |
| 3. Moderate Model Pluralism | There exists an aspect $x$ of a phenomenon $y$ such that scientists need multiple models to explain/predict $x$. |
| 4. Strong Model Pluralism | For almost any aspect $x$ of phenomenon $y$, scientists require multiple models to achieve scientific goal $z$. |

**Table 1:** Four Kinds of Model Pluralism

Table 1 only offers a rough sketch, but it provides an elegant summary of the structure and content for the following sub-sections, each of which will address one of the four kinds of model pluralism. In the subsequent discussion, I shall argue that model pluralism in its strong form is to be embraced.

### 2.1 Weak Model Pluralism
I consider weak model pluralism as the received view in the philosophy of science. Indeed, almost everyone in the field recognizes that the phenomena under scientific investigation are incredibly diverse, and scientists need different models to explain/predict these different phenomena. Hardly worthy of the label model pluralism, this view is sometimes minimally expressed in the recognition that different scientific disciplines have and require different models.[11] Indeed, let us quickly move on to a slightly more 'radical' position.

### 2.2 Weakly Moderate Model Pluralism
The core of weakly moderate model pluralism can be defined as follows: each phenomenon has many different aspects, and scientists need different models to explain/predict these different aspects of a single phenomenon. This view, I gather, is emerging as the new dominant view in the philosophy of science. Much of this new consensus comes out of work in the philosophy of economics and philosophy of biology. Indeed, in order to demonstrate the need for more pluralist views on MMM, it will prove very useful to look at the brief history of philosophy of economics, a history that has primarily focused on the scientific status of highly abstract and idealized models.

---
[11] A better label would be Relaxed Model Monism.

Unlike any other discipline, economics has been faced with suspicion and criticism, from both within academia and outside of it. The opposition to economics peaked with its 'failure' to predict the global financial crisis of 2008. Thereafter, even many economists began to challenge the standards and methods of the discipline, demanding a change for the better. Many deduced that the problem of economics is grounded in its models. Criticism has been particularly directed against a variety of features economic models seem to share.

Consider the following popular statements among critics: (1) economic models are unrealistic, (2) economics is too simple, (3) economics leaves out important features of the real world, and perhaps the apparently most damning feature of all (4) economic models are based on false assumptions. Many more could be named, but they seem to share a common theme that is directed against the *simplicity* of these models.

Economists, however, seemingly unaffected by the financial crisis, have largely ignored the criticism directed against the methodology of their discipline. Despite the 'meant to be' damning criticism, economists continue to show confidence in their models and their ability to explain and predict real-world phenomena. Of course, no economist would deny that models in economics are highly abstract, idealized and simple nor that complex or very realistic models are rare if not the exception. This, I believe, suggests a disciplinary standard within economics, that equates good models with very simple ones. If simple models are the norm, unrealisticness becomes a straightforward result of idealization and abstraction.

Aydinonat (2018a) has suggested a different, albeit illuminating reading. He suggests that criticism is being directed against the unrealisticness of economics models. These two readings are of course not exclusive, but I suggest that the source of criticism against the unrealisticness of economics is primarily grounded in the simplicity of economic models. The inverse, i.e. simple models being built in order to achieve unrealistic models of the real world, seems to obscure what economists are actually interested in. Nevertheless, charitably interpreted such criticism may entail an important revelation. If simplicity is an overvalued standard of economic modeling, the result may very well be the accumulation of models that fail to explain or predict, let alone depict an accurate representation of the world. This is a real danger.

After all, even if the official subject of a discipline concerns the provision of successful explanations and predictions in a specific domain, norms within the community of a discipline may very well be able to undermine such 'official goals'. The obvious example here is the recent controversy surrounding the replication crisis and the use of statistics within science. Amrhein, Trafimow & Greenland (2019) published a highly influential article, in which they argued that the concept of statistical significance ought to be discarded.[12]

> The empire of "statistical significance" has its roots in the 19th-century writings of Edgeworth (1885) and reached full dominance with the spread of cutoffs for testing, formalized by Jerzy Neyman and Egon Pearson as Type-I error rates. Like the political empires of their period, such hypothesis testing for scientific (as opposed to mechanical) inference is a relic of a bygone era, whose destructive effects reverberate to this day. We hope this era is over. (Amrhein, Trafimow & Greenland 2019, p. 266)

If journals - and by extensions reviewers and editors - are reluctant to publish results failing to meet the standards of an arbitrarily chosen but entrenched significance level, a bias is created that has negative effects on the progress and the understanding of science itself. After all, statistical significance does not imply scientific significance. Nevertheless, the methodological problems with statistical significance testing deserve proper philosophical attention themselves, attention which I cannot provide here due to the scope of the paper. However, this example should illustrate how misguided norms within science can have a destructive effect on a discipline.[13]

---

[12] In fact, their criticism of misguided norms concerning statistical testing has gathered over 800 signatures of support by researchers across the globe in a mere week.

As a result, they conclude that we "should treat statistical results as being much more incomplete and uncertain than is currently the norm" (2019, p. 262). A similar thought, I suggest, is at work in the now vast philosophical literature on modeling. Many philosophers argue that highly abstract and idealized models are incomplete and should be treated as far less certain than scientists often take them to be.[14] This criticism is of a general methodological sort. It is a philosophy of science issue concerning the very nature of modeling. Hence, these conclusions not only apply to economics, but also to other sciences such as physics and biology. But as economics, unlike other disciplines, relies almost exclusively on such simple models, these methodological concerns hit much harder than for other sciences that employ a much wider range of experimental methods, complex models and simulations. Indeed, some philosophers have gone so far as to criticize economists for engaging in mere mathematical fancy, not engaging with the empirical world, and in fact, not even conducting a science (see Rosenberg 1983). Despite the increasing popularity of behavioural economics, models within economics have remained highly abstract and idealized. This has puzzled many economic methodologists and philosophers of economics. In fact, the majority of work within philosophy of economics seems to be 'obsessed' with the scientific status of such simple models. This obsession, however, is unsurprising. Whatever consensus is about to emerge on the status of models in economics, it is obvious that the answers provided will have considerable downstream effects on the status of economics itself. The possible results range from a complete validation of current economic practice to a complete dismissal of economic modeling; instead validating the general distrust in economics. The stakes are high.

Progress in our understanding of models has largely recognized that economics does not proceed as envisioned by either economists or their 'naïve' critics. Despite some philosophers highlighting the premature criticisms of simple models, noting the diverse and important roles such models play, they nevertheless agree that many scientists treat single models with too much confidence.[15] Economists, however, as previously alluded to have largely ignored this literature.[16] Dani Rodrik (2018) suggests two reasons beyond mere disinterest: economic methodology is "relegated to specialized journals" and, more importantly, is not really being taught in grad school (p. 276). Nevertheless, some economists have recently engaged with philosophical questions themselves, criticizing the tendency in economics to ignore methodological concerns, and seeking contact with philosophers of economics in the process.[17] One work that stands out, in particular, is Rodrik's (2015) recent book *Economics Rules. Why Economics Works, When It Fails, and How to Tell the Difference.* Rodrik, a Harvard professor of economics, provides a sincere attempt to bridge the gap between economic practitioners and critics of economics, providing a "practitioners' attempt to explain (to himself and others) how economics works, when it does, and how it fails, as it often does" (Rodrik 2018, 276) that both "celebrates and critiques economics" (Rodrik 2015, p. 6). He argues that models "are both economics' strength and its Achilles' heel" (2015, p. 5) supporting my earlier suggestion that economics stands or falls by the explanatory and predictive power simple models can provide. In trying to understand the role of models in economics, Rodrik, drawing on the work of philosophers of economics, argues that while economic models do have limits, these limits can be overcome by embracing what I have called *model diversity*. The philosopher of economics N. Emrah Aydinonat (2018b) calls this the "motto of the book" concluding from Rodrik's analysis that the "diversity of models is a means to better explanations" (p. 237). It is not only a descriptive claim about science, i.e. that model-based science operates by the creation of diverse models, but also a normative one: diversity of models is a strength rather than a weakness and ought to be sought in order to achieve scientific progress.

---

[13] There might be an important relation here between the norms within scientific disciplines and Thomas Kuhn's (1962) idea of scientific research paradigms, an idea I can unfortunately not explore here due to considerations of space.
[14] See Cartwright (1983, 2009); Hausman (1992); Mäki (2009); de Donato & Zamora-Bonilla (2009); Grüne-Yanoff (2009); Knuuttila (2009); Kuorikoski & Lehtinen (2009); Northcott & Alexandrova (2015); Fumagalli (2016); Marchionni (2017).
[15] See Wimsatt 2007; Weisberg (2007), (2013); Ylikoski & Aydinonat 2014; Aydinonat (2018).
[16] To their defense, so have most scientists.
[17] See Sugden (2000, 2001, 2009, 2011, 2013); Gilboa, Postlewaite, Samuelson, & Schmeidler (2014); Rodrik (2015, 2018).

Indeed, Rodrik is to some degree an outlier among economists, a state he recognizes himself. Though often a fervent critic of his fellow economists, he "felt that many of the criticisms coming from outside the field missed the point" (2015, p. ix). Rodrik argues that economists themselves are to blame for the misinformed picture of economics found in the public. Many economists, so he argues, take economic models too literally "pronouncing universal economic laws that hold everywhere, regardless of context" (p. x). His book, therefore, has two audiences in mind: both economists and non-economists interested in economics (in particular critics). Economists, he argues, are simply bad at conveying how their own discipline works, perhaps due to a lack of interest in, and only a shallow understanding of the methodological questions economics raises. However, unlike much of the literature in the philosophy of economics, Rodrik does not argue for a change of economics itself. It is just that economists "need a better story about the kind of science they practice" (p. x). This is an interesting turn in a literature that has often tried to give prescriptions on how economics ought to be done if it is to be recognized as a science (see Rosenberg 1976; 1983; 2009). Rodrik offers an alternative descriptive picture of economics that does not exhibit many of the flaws economics seems to have, in virtue of how economists represent it publicly. This offers a much more positive picture of economics that Rodrik carefully suggests should not only be appreciated but could serve as a role model for the improvement of the other social sciences. In fact, I shall argue that an embracement of model pluralism might be necessary for the study of all complex systems, including physics, chemistry and biology. The key to this alternative description of economics lies as Aydinonat (2018b) points out, in one particular sentence in Rodrik's book: "It's *a* model, not *the* model" (2015, p. 43). Rodrik (2018) suggests the following mantra within economics that serves as a standard reply when economic models are criticized:

> A model is an abstract, simplified setup that sheds light on the economy's workings, by clarifying the relationship among exogenous determinants, endogenous effects, and intermediating processes. Economic science advances by testing these models against reality, keeping those that do a good job and discarding the rest. (2018, p. 276)

The second sentence in this standard reply he believes is to blame for much of economics' failures. Economists seem to believe that their science operates according to a simple form of falsificationism, due to Popper (1959; 1963), a picture of science that, most contemporary philosophers of science would agree, both misrepresents scientific practice and provides the wrong ideal of how science ought to function.[18] While Popper was right to suggest that the creation of hypotheses - and let us extend this to models - is very much a craft and a creative process, models in economics are rarely if ever discarded or replaced by better ones. This is not a mere artefact of the lack of data or inadequate empirical methods, but rather, so Rodrik (2018) suggests, an unavoidable feature of the social sciences:

> [r]ather than a single, specific model, economics encompasses a collection of models. The discipline advances by expanding its library of models and by improving the mapping between these models and the real world. The diversity of models in economics is the necessary counterpart to the flexibility of the social world. Different social settings require different models. Economists are unlikely ever to uncover universal, general purpose models. (Rodrik 2015, p. 5)

---

[18] Due to the scope of the paper, I can unfortunately not offer a sustained attack on falsificationism, but can point to a number of highly influential criticisms that have undermined the popularity of falsificationism among philosophers of science. See Kuhn (1962); Lakatos (1970, 1980); Putnam (1975), Feyerabend (1975), Ruse (1977), Kitcher (1982), and Hacking (1983).

Rodrik argues that this picture has not caught on among economists due to a sort of physics envy that has also been attested by a number of other philosophers (see Rosenberg 2009; Sugden 2000; 2001; 2009), leading economists to "misuse their models" (Rodrik 2015, p. 5). Economists, he argues, must overcome their temptation to generalize the results of singular models mistaking "*a* model for *the* model, relevant and applicable under all conditions" (p. 6).[19] This insight, however is far from a negative one. Instead, he argues that "each economic model is like a partial map that illuminates a fragment of the terrain" jointly serving as "our best cognitive guide to the endless hills and valleys that constitute social experience" (2015, p. 8). This picture is an illuminating one. Once model pluralism is embraced as a more accurate description of actual economic practice, "standard criticisms of economics lose their bite under this alternative account" (2015, p. x). Rodrik's analysis suggests that economic models must appear rather weak in isolation when the goal is to explain phenomena.[20] Rather than providing highly complex models that are hardly tractable or useful, economists, he argues, provide sets of simple and abstract models, each focusing on different what-if questions to analyse different aspects of certain phenomena. While Rodrik's (2015) plea for model diversity is quite novel in the literature on economic methodology, it is a far cry to call it a completely new idea in the philosophy of science. Widening our perspective across the disciplinary gaps of economics it must become obvious that the biologist Richard Levins, in an article as old as 1966, argued for an analogous position in the case of biology.

> The multiplicity of models is imposed by the contradictory demands of a complex, heterogeneous nature and a mind that can only cope with few variables at a time; by the contradictory desiderata of generality, realism, and precision; by the need to understand and also to control; even by the aesthetic standards which emphasize the stark simplicity and power of a general theorem as against the richness and the diversity of living nature. These conflicts are irreconcilable. Therefore, the alternative approaches even of contending schools are part of a larger mixed strategy. But the conflict is about method, not nature, for the individual models, while they are essential for understanding reality, should not be confused with that reality itself. (1966, p. 431)

Levins argued that among multiple goals one may have in the creation of a model, himself focusing on generality, realism and precision, only two can be maximized. Akin to Rodrik (2015), Levins (1966) argued against naive versions of model monism. Philosophers, such as Aydinonat (2018a, 2018b), Weisberg (2007, 2013) and Godfrey-Smith (2006) have argued for a more profound message in Rodriks' and Levins' respective works that I shall label moderate model pluralism.

## 2.3 Moderate Model Pluralism

Moderate model pluralism follows straightforwardly from the Levins' and Rodrik's arguments for weakly moderate model pluralism. It is the thesis that there are aspects of particular phenomena such that scientists require multiple models in order to explain or predict. Drawing on Levins (1966), Weisberg (2013) argues that inherent trade-offs between the diverse modeling goals are inherent *features* of model-building. Hence, there cannot be an all-purpose model, perhaps not even for one specific research question. Indeed, Weisberg has done much to bring an awareness for trade-offs into the literature on MMM. Retrospectively, it may appear obvious that Weisberg is right, but it took much effort and intricate analysis to demonstrate that model monism is simply untenable. In that, my model pluralism account owes much to Weisberg's previous work. Indeed, he (2007, 2013) further suggests that a strategy he dubs *multiple model idealization* - though, hitherto not having received much philosophical attention - provides a better picture of model-based science.[21] This insight is important. Given the plurality of scientific desiderata, the constraints "imposed by logic, mathematics, and the nature of representation, conspire against" their simultaneous achievement in a single model (Weisberg 2013, p. 104). Weisberg's (2013) monograph provides a superior picture to the received monist views of model-based science. However, he needlessly restricts himself to define multiple model idealization as the "practice of building multiple related but incompatible models, each of which makes distinct claims about the nature and causal structure giving rise to a phenomenon" (2007, p. 645). Modelers, after all, often create compatible models in order to increase our confidence that a process akin to the one in the model could be at work in the real world, something Weisberg clearly recognizes when he defends robustness analysis as a sort of low-level confirmation.

---

[19] A similar criticism of physics been provided in Nancy Cartwright's (1983) *How the Laws of Physics Lie*.
[20] See also Ylikoski and Aydinonat (2014).

The epistemic strength of multiple models, however, goes beyond mere robustness analysis and is the core claim of my model pluralism account. The latter insight is what I take to be the takeaway Aydinonat (2018b) draws from Harvard economist Dani Rodrik's (2015) recent call for model diversity in economics, with sets of simple models possibly outweighing the explanatory power of particular complex and more realistic models. If one recognizes that the plurality of models is a necessary and ubiquitous feature of science, Potochnik (2017) is right to point out that the "category of multiple-models idealization" becomes "so broad that it is a sort of dustbin category, uninformative about the features of idealizations that fall into it" (p. 46), indeed it should not even be considered a particular form of idealization to be contrasted with Weisberg's (2007) alternatives *Galilean* and *minimal idealization*. Model pluralism is a much broader idea than the narrow case of multiple model idealization suggested by Weisberg. Indeed, Potochnik's worry applies more generally. There is no univocal account of the epistemic roles multiple models play without sacrificing all informativeness. Unlike Weisberg's taxonomy of different kinds of idealization suggests, science almost always requires the use of multiple models. What may appear to be Galilean or minimal idealization, in fact, involves a variety of background models that easily missed out by moving away from the scientific modeling practice. This is the core argument for strong model pluralism.

Nevertheless, Weisberg (2007) mentions further pragmatic and explanatory reasons for using multiple models, all of which support strong model pluralism more generally. First, multiple models can serve the maximization of predictive power in the case of weather forecasts. One model might be better than the others, but the best predictions will be provided by relying on the whole set of available models. If one out of a set of models predicts catastrophic environmental effects, for instance, we should not disregard the information gained by the model merely because there are one or more models that have on average provided superior predictions. Clearly, the special roles of models within policy-making would support model diversity, even when the scientists within the field are committed to the idea that there is something like the 'best' model. Similarly, a diversity of economic models may vastly improve our ability to predict economic crises. Second, as pointed out by population biologists such as Roughgarden (1979) and May (2001), "clusters of simple models increase the generality of a theoretical framework, which can lead to greater explanatory depth" (Weisberg 2007, p. 647). Akin to the vast number of models created via robustness analysis, model diversity enables a much deeper understanding of the causal factors at play than one or a few simple models would have generated. Further, Weisberg (2007) argues that such sets of simple models can be used for the creation of new structures, mentioning engineering and synthetic chemistry. Similarly, economic models, in particular game theoretic ones, have served in the construction of auctions, optimizing the payoffs for governments. Finally, Weisberg suggests that "building a set of models that gives maximum generality, at the expense of capturing all of the core causal factors" (p. 648) can provide general explanation schemas.

---

[21] Especially when it comes to modeling complex phenomena (see also Wimsatt 2007).

Having rehashed Weisberg's arguments for model diversity it is somewhat surprising how little attention he pays to multiple-model idealization. Within the existing literature, however, this is unsurprising. Weisberg has done more than anyone to move us away from focusing on single models, and his contribution must be seen as a substantial improvement within the existing (and largely monist) literature.

However, I think more radical conclusions follow from Weisberg's own arguments for model diversity than he himself anticipated. Let me suggest an analogy, that draws on the often-made claim that modeling is a craft or art.[22] Imagine the average handyman, faced with the task of repairing a sink. In order to achieve his goals, he requires multiple tools, none of which was specifically designed for sink repair. The procedure requires multiple steps, each of which could be satisfied by a variety of tools. While one workman chooses a set of tools including tool a, b, and c, another might have chosen tool x, y, and z. At no step in this procedure would they think that the tools they did not choose should be discarded. Having a diverse set of tools is a benefit, not a problem, and we would be well advised to extend our set of available epistemic tools. If anything, model diversity is a source of strength calling for the use of multiple models when there is no clear indication of which models would serve a specific purpose better. Unlike a handyman moving from town to town, a scientist is not limited by the model he carries along, but rather by his cognitive capacities to use further models. Science, however, is a collective discipline. If different scientists use different models and become proficient in different modeling practices to explore the same questions, we should not see this as a bug that needs fixing but rather an indication of the richness and diversity the discipline has achieved. Model pluralism is an inherently pragmatist and context-sensitive position. The following section on strong model pluralism will illustrate that model monists can embrace a radically pragmatic and context-sensitive position and still misrepresent scientific modeling practice and the epistemic roles of models by focusing on single rather than multiple models.

## 2.4 Strong Model Pluralism

In light of the foregoing arguments moderate model pluralism does not appear in any way to be controversial position. What emerges instead, is a recognition that philosophers *hitherto* have been exceedingly critical of such confidence in highly idealized models by solely focusing on the relationship between a single model and the real world. Model pluralism, on the other hand, fully embraces this diversity of models and modeling-practices in the scientific toolkit. Accordingly, I will now explicate and defend the following two core theses of model pluralism: (i) any successful analysis of models must target sets of models, their multiplicity of functions within science, and their scientific context and history and (ii) for almost any aspect *x* of phenomenon *y*, scientists require multiple models to achieve scientific goal *z*.

---

[22] See Feyerabend (1975) for a general case of science as an art, and Rodrik (2015) for a case-study of the importance creativity and diversity plays in economic modeling.

Here, somewhat ironically, a famous model from the social sciences serves as an elegant case to contrast my model pluralism account with the aforementioned philosophical literature. As illustrated by the vast literature on models, many of the papers written in this debate focus on a particular philosophical question, e.g. "how can highly abstract and idealized models explain?", and attempts to answer this question by analysing one model that is recognized as successful, or at least, highly influential by the scientific community. The most discussed and cited model in the philosophical literature on models is of course the 'Sakoda-Schelling model of segregation', usually referred to as the 'Schelling model', the 'checkerboard model', or the 'checkerboard model for racial segregation'.[23] Much of the apparent staleness of the literature may simply come from the observation that *yet another* paper has been published, discussing *this* rather than *any other* model. Indeed, the list of papers analysing the Sakoda-Schelling model is vast, see for instance: Sugden (2000), Aydinonat (2007; 2008), Clark and Fossett (2008), Weisberg (2013), Ylikoski and Aydinonat (2014), Fumagalli (2016), Lisciandra (2017), and Verreault-Julien (2019). Naturally, this is just one instance where monist thinking had a negative influence on the literature, motivating Weisberg (2013) to focus on a variety of very different models, i.e. in addition to the Sakdoda-Schelling model.

Indeed, *the* checkerboard model is often treated as an ideal paradigmatic model from which inferences to other models are to be drawn. Rather than treating it as a mere *token*, it is treated as an *ideal instance of a type*. This tendency to generalize from particular models is to be avoided and my model pluralism account explains why. Nevertheless, the extraordinary attention the Sakoda-Schelling model has received is unsurprising, after all, as one of the earliest agent-based models it has only recently been overtaken as the most cited article in the Journal of Mathematical Sociology, making it one of the most influential models in the social sciences.[24] Further, it is an incredibly simple model, requiring no knowledge of physics, chemistry, biology, or the social sciences, for that matter. Hence, it serves as a highly attractive target of philosophical investigation for philosophers of science regardless of their specialization.

In the philosophical literature on models, Ylikoski and Aydinonat (2014) have made a very important observation here that has not yet received the attention in the literature it should have. Shedding light on the discrepancy between the philosophy of models literature and actual scientific modeling practice, they point out that although the 'Schelling model' is often treated as a single model by philosophers, the name does neither refer to Schelling's original model nor Schelling's original set of models. Instead, they argue the 'Schelling model' refers to a "whole cluster of models that are related to each other through genealogical origin and similarity" (p. 22), calling this the *family of models thesis*. This 'family' will continuously grow, with variations of the original 'Schelling model' appearing on the scientific landscape.

To some extent, this broader idea of understanding the epistemic contribution of models in virtue of their 'offspring' models, is already well recognized in the philosophical literature on robustness analysis, i.e. testing the robustness of a models results via the introduction of minor changes to the model, resulting in a broad set of models.[25] Their *family of models thesis*, goes beyond robustness analysis, however, with some members of the set only related to the original model through several steps removed. After several slight alterations of a model, one could almost call them mutations, the resulting sets of models is hardly the mere product of robustness analysis. Even a broad sense of robustness analysis is not going to cover the resulting diversity that is highly distinct from the original model; neither would it correspond to the way the term 'robustness analysis' is used among modelers. Nevertheless, as Chiara Lisciandra (2017) points out:

> the problem of how to compare results deriving from structurally different models is one of the most interesting questions that the debate on robustness analysis has opened to today's scientific practice and promising works are expected to come from this research area in the near future" (p. 83).

---

[23] Rainer Hegselmann (2017) in a lengthy and through treatment of Thomas C. Schelling (1971, 1978) and James Sakoda (1971), provided a historical analysis and proof of mathematical equivalence showing that James Sakoda (1971), though an unfortunate victim of the Matthew effect, is the true inventor of the 'checkerboard model' first developing the idea in his 1949 dissertation.

[24] Cf. the JMS website at http://www.tandfonline.com/toc/gmas20/current, while in 2017, Hegselmann (2017) reports it as the most cited article in the journal.

[25] See Lisciandra (2017) for a recent overview, but also Woodward 2006.

Not only is it possible to usefully think about such smaller sets of models, it is also possible to consider a larger set of models than the *family of models thesis* suggests. While Ylikoski and Aydinonat (2014) insist on a rather strict form of both similarity and offspring relationships, this is not necessary. First, it could be useful to analyse incredibly dissimilar set of models, that nevertheless stand in in a genealogical relationship. Ylikoski and Aydinonat make a substantial contribution to the literature by leading us away from the simple monist appeal of trying to analyse the epistemic contribution of single models without the consideration of the larger context in which the model operates. However, they do not go far enough and fall back into the trap of monist thinking by trying to tie the semantics of what modelers mean when they speak of 'the Schelling model' to the epistemic contribution of the model. As Weisberg (2013) himself points out, there are different levels of analysis concerning models. We can take a sociological-historical approach, an epistemological approach, or a metaphysical one. While at the heart of much of the MMM literature, the latter is an avenue without much promise of success.[26] Any successful analysis of models must target sets of models, their multiplicity of functions within science, and their scientific context and history.

As such, it might be useful to idealize the family of models hypothesis towards a *population of models hypothesis*. This shifts the focus away from the semantics of what scientists mean when they refer to the success of a particular model, and towards a pragmatically justified conceptual scheme. Such a scheme is pragmatic and context-sensitive. Such a scheme would be less permissive towards the misunderstanding of the contributions particular models made. I am, however, somewhat reluctant to use of the term *population of models*, as it still may lead some to mistakenly restrict themselves to treating models as only jointly explanatory if they have common ancestry. Rather it should be treated as a useful conceptual tool to think about particular models and their epistemic contributions. As Hegselmann (2017) illustrates, Sakoda (1971) invented the checkerboard model before Schelling (1971). But even without a genealogical relationship between their models, they are clearly very similar, as Hegselmann's proof of mathematical equivalence emphasizes. While an exclusion of the Sakoda-model and its 'ancestors' might be particularly useful for the purposes of sociological or historical analysis, grouping them together within a set of models could serve a multiplicity of important epistemic purposes.[27]

Hence, a stronger conclusion emerges from Rodrik's (2015) description of economics. It is only together that models can illuminate the mechanisms that are truly at work in complex target systems. However, though Aydinonat (2018b) argues that his analysis only provides a descriptive account of actual economic modeling practice, it suggests an explanatory factor that has not received much philosophical scrutiny but could partly explain the confidence in scientific models across disciplines. However, I am not much interested here in the specifics of Rodrik's (2015) complete account if one could even call it that. After all, it is not meant as a "treatise on economic methodology" (Rodrik 2018, p. 276), and hence unsurprisingly faces several methodological problems once it is treated as one.[28] Instead, I am more interested in the general idea of model pluralism with a much stronger conclusion than Rodrik's call for model diversity. Rodrik (2018) may be justified in invoking Ockham's razor to "use the least number of models as possible" (p. 278), but in almost all cases this will involve more models than an abstract arm-chair analysis suggests. Even when scientists specifically talk about one particular model, they will implicitly have background models in mind that are invisible to those not embedded within the scientific practice of modelers. Philosophers who attempt to understand model-based science by analysing particular models instead of sets of models commit a fatal mistake.[29]

---

[26] See also O'Connor and Weatherall (2016).
[27] The common tendency to rationalize a single model in virtue of its success should be avoided (see Veit, Dewhurst, Dolega, Jones, Stanley, Frankish, and Dennett on "The Rationale of Rationalization", forthcoming).

## 3 Conclusion

In this paper, I have defended an account of models, modeling, and model-based science that takes plurality, diversity, and richness of MMM seriously, hence dubbing it *model pluralism*. In explicating this account, I have defended the following two core theses: (i) any successful analysis of models must target sets of models, their multiplicity of functions within science, and their scientific context and history and (ii) for almost any aspect $x$ of phenomenon y, scientists require multiple models to achieve scientific goal $z$. Neither of these conclusions has yet been recognized in the MMM literature. Indeed, the foregoing discussion allows us to draw a continuum of 'pluralist' thinkers within the debate:

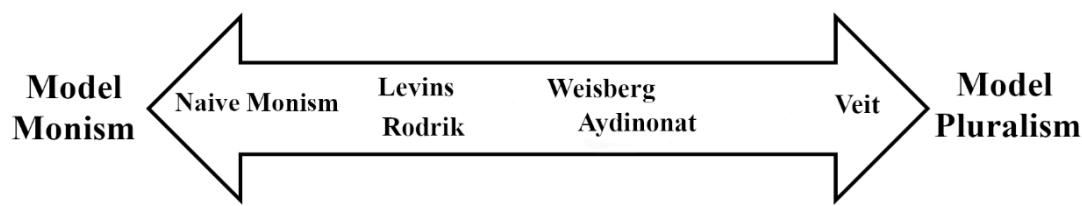

**Figure 1** The Pluralist Continuum

Weisberg, Aydinonat, and their scientific inspirations, i.e. Levins, and Rodrik respectively, have provided important arguments that have served as stepping stones for the more thoroughgoing pluralism presented here (Figure 1). Indeed, it is only the 'strong model pluralism' defended here that deserves the label model pluralism. Weisberg's (2007, 2013) and Aydinonat's (2018b) accounts, despite seeming somewhat radical against a background of more monist accounts, are actually quite moderate positions that attempt to accommodate monist motivations. Model pluralism entirely relinquishes the essential tenet of model monism to understand models by analysing single models. As I have argued in this paper, model pluralism is nothing to be afraid of. It is a general account of models, modeling, and model-based science. In analogy to the to problematic norms of statistical significance I alluded to in the introduction, Amrhein, Trafimow & Greenland (2019) provide a conclusion similar to the one I draw here on model pluralism:

> for what comes next, there is no substitute for accepting methodologic diversity (Good 1957; Cox 1978; Box 1980; Barnard 1996; Little 2006; Senn 2011; Efron and Hastie 2016; Crane

---

[28] See Mäki (2018), Grüne-Yanoff and Marchionni (2018) and Aydinonat (2018b), for an illumination of these difficulties.

[29] In Veit (forthcoming) I discuss this problem in relation to the role and impact of models in the scientific study of morality. Single models of cooperation, so argued by many can neither explain morality, nor explain it away (see Veit 2019b). In order to understand the significance of these models, however, they need to be considered jointly. Similarly, Veit (2019a) discusses the scientific virtues of considering alternative models in the Major Transitions of Evolution.

2017), with careful assessment of uncertainty as the core motivation for statistical practice […]. (Amrhein, Trafimow & Greenland 2019, p. 266)

Once model pluralism is taken seriously there is, in fact, no substitute for accepting methodological diversity. The epistemic roles of particular models can only be understood against their scientific context and history, often including quite large sets of other models. One may even alternatively label it model holism. Far from over, philosophers working on model**s** have a behemoth of work in front of them, with economics providing an elegant field of investigation. The widespread criticism of economics has led to the discovery of an unavoidable feature of science. If what I suggested in this paper is correct, economics will serve as a highly attractive field of investigation for philosophers of science. After all, it very much stands or falls with the strength of model pluralism. Rather than showing that economics is an inherently flawed discipline, philosophers and methodologists of economics may have found the very key to understand the success of science. Hence, I suggest that philosophers of science should turn more of their attention towards both the social sciences and the philosophy of the social sciences, a historical omission that if my argument is correct and model pluralism is unavoidable, may have provided a misguided understanding of how science works. To conclude: model diversity is a feature, not a bug. In order to understand science, the philosophical analysis of models, modeling, and model-based science must focus on sets of models, their multiplicity of functions within science, and their scientific context and history.

**Acknowledgments**
First of all, I would like to thank Rainer Hegselmann, Paul Teller, Cailin O'Connor, Robert Sugden, Caterina Marchionni, Donal Khosrowi, Lena Zuchowski, Joe Dewhurst, Shaun Stanley, Heather Browning, and two anonymous referees for comments on different stages of this paper. Secondly, I would like to thank audiences at the 11$^{th}$ MuST Conference in Philosophy of Science at the University of Turin, 2018's Model-Based Reasoning Conference at the University of Seville, the 3rd think! Conference at the University of Bayreuth (especially Johanna Thoma), the 4th FINO Graduate Conference in Vercelli, the Third International Conference of the German Society for Philosophy of Science at the University of Cologne, the PG WiP at the University of Bristol (in particular Aadil Kurji), the 26th Conference of the European Society for Philosophy and Psychology at the University of Rijeka, the Lake Como summer school for Economic Behavior, the 16$^{th}$ Congress on Logic, Methodology, and Philosophy of Science and Technology in Prague, the 14$^{th}$ INEM conference at the University of Helsinki (in particular Till Grüne-Yanoff and Chiara Lisciandra), the Philosophy of Social Science Roundtable at the University of Vermont, and an invited lecture at the Centre for Networks and Collective Behaviour [University of Bath]. Sincere apologies to anyone I forgot to mention.

**Declaration of Conflicting Interests**
The author(s) declared no potential conflicts of interest with respect to the research, authorship, and/or publication of this article.

**Funding**
The author(s) received no financial support for the research, authorship, and/or publication of this article.

**Author Biography**

Walter Veit is a Pre-Doctoral Research Fellow at Carnegie Mellon University. His main research is on cognitive science, theoretical biology, economic methodology, bioethics, and the philosophy of science.